# Phase-field simulation of domain switching in ferroelectric trilayer films under bending-induced strain gradient[*]

GUO Changqing[1, 2], YANG Letao[1, 2], WANG Jing[1, 2], HUANG Houbing[1, 2, *]

1. School of Materials Science and Engineering, Beijing Institute of Technology, Beijing 100081, China
2. Advanced Research Institute of Multidisciplinary Sciences, Beijing Institute of Technology, Beijing 100081, China

**Abstract**

Flexible ferroelectric materials possess considerable potential for wearable electronics and bio-inspired devices; however, their electromechanical coupling mechanisms under dynamic bending conditions are not yet fully understood. In this work, the effects of bending deformation on domain structures and macroscopic ferroelectric responses in $(SrTiO_3)_{10}/(PbTiO_3)_{10}/(SrTiO_3)_{10}$ flexible ferroelectric trilayer films are systematically investigated using phase-field simulations. By constructing computational models for upward-concave (U-shaped) and downward-concave (N-shaped) bending configurations, the strain distribution and its regulation mechanism on polarization patterns under different curvature radii are analyzed. The results show that the two bending modes generate completely opposite strain gradients along the film thickness: U-shaped bending produces compressive strain in the upper layer and tensile strain in the lower layer, whereas N-shaped bending yields the reverse distribution. Such inhomogeneous strains drive significant polarization reconfiguration within the PTO layer. At a moderate curvature (large $R$), the system retains stable vortex-antivortex pairs. Reducing bending radius (smaller $R$) promotes divergent topological transitions—U-shaped bending facilitates vortex pair transformation into zigzag-like domains, while N-shaped bending drives vortex-to-out-of-plane $c$-domain evolution. Notably, bending-induced strain gradients impose transverse flexoelectric fields that markedly change trilayer hysteresis loops. U-shaped bending introduces a negative flexoelectric field, shifting loops rightward with maximum polarization ($P_{max}$) decreasing. In contrast, N-shaped bending generates a positive flexoelectric field, enhancing $P_{max}$ via leftward loop shifting. Analysis of polarization switching under applied electric fields further demonstrates that bending mediates the control of domain-evolution pathways and reversal dynamics. These findings not only elucidate profound bending effects on flexible ferroelectrics' domain architectures and functional properties but also provide theoretical guidance for designing strain-programmable ferroelectric memories, adaptive sensors, and neuromorphic electronics.

# 1. Introduction

Ferroelectric perovskite oxides have become core functional materials in modern electronic devices due to their unique spontaneous polarization and their dynamic responsiveness to external electric and mechanical fields [1,2]. These materials not only exhibit excellent ferroelectric, piezoelectric and dielectric properties, but also have been widely used in non-volatile memory[3,4], high-precision piezoelectric sensors[5-7], high-frequency filters[8] and other key devices owing to their high Curie temperatures, robust polarization stability, and low dielectric loss. In addition, they also exhibit great application potential in medical ultrasound imaging, precision electromechanical systems, and microelectromechanical actuators and other fields [9-11].

With the rapid development of flexible electronic technology, such as wearable health monitoring devices, artificial electronic skin, bionic soft robots and so on, ferroelectric materials with both high electronic performance and mechanical flexibility are facing new challenges[12-15]. However, traditional inorganic ferroelectric oxides exhibit high intrinsic brittleness due to their strong ionic or covalent bond-dominated crystal structure. In addition, the mechanical clamping imposed by rigid substrates further suppresses free domain evolution, thereby reducing polarization-switching efficiency and weakening the dynamic response of the hysteresis loop to applied fields [16,17]. These factors significantly limit the application of traditional ferroelectric materials in flexible electronics. Therefore, a key scientific question for the development of flexible ferroelectric devices is how to enhance mechanical flexibility while preserving excellent ferroelectric properties. In recent years, advances in thin-film fabrication techniques have opened new pathways for producing high-quality freestanding ferroelectric films with excellent mechanical flexibility, elasticity, and electrical performance. For example, freestanding single-crystal ferroelectric thin film [18-23] such as BaTiO$_3$ and BiFeO$_3$ have been successfully fabricated using pulsed laser deposition (PLD) combined with sacrificial-layer etching techniques, including the use of a water-soluble Sr$_3$Al$_2$O$_6$ layer. These films do not fracture during 180 ° folding and can return to their original morphology after removal of the bending load, thereby overcoming the constraint of a rigid substrate. More importantly, benefiting from strain engineering and strain-gradient engineering, these freestanding ferroelectric thin films exhibit enhanced tunability in their polarization distributions and topological domain structures. In this context, (PbTiO$_3$)$_n$/(SrTiO$_3$)$_n$ superlattice system have emerged as an important research platform in ferroelectric studies owing to their atomic-level heterointerface engineering capabilities. The synergistic effect of interfacial strain, electrostatic coupling and gradient energy in this system enables it to stabilize a variety of nontrivial ferroelectric domain structures, including flux-closure domains[24], polar vortices[25-27], and skyrmions [28]. For example, Yadav et al. [25] observed periodic array of vortex-antivortex arrays in (SrTiO$_3$)$_{10}$/ (PbTiO$_3$)$_{10}$ superlattices by scanning transmission electron microscopy, in which the polarization vector rotates continuously along closed trajectories. The formation of these topological domains arises from a dynamic balance among competing energy contributions, including the elastic energy induced by lattice mismatch, the electrostatic energy associated with polarization discontinuity between the PbTiO$_3$ and SrTiO$_3$ layers, and the gradient energy required for polarization rotation or amplitude variation. Moreover, the periodic rotation of polarization further reduces the total free energy of the system.

In ferroelectric thin-film research, strain engineering has been widely employed to modulate lattice distortions, thereby influencing the stability of domain structures and the kinetics of polarization switching. According to classical thermodynamic theory, in-plane compressive strain can suppress the formation of in-plane $a$ domains in PbTiO$_3$ and enhance the stability of out-of-plane $c$ domains, thereby increasing the remanent polarization. However, in the actual working environment of flexible devices, the film is often in a non-uniform strain state, resulting in a significant strain gradient in the material, which leads to the flexoelectric effect [29]. Unlike uniform strain, strain gradient not only changes the local polarization direction, but also may affect the evolution dynamics of domain walls. Although the effect of strain gradient on ferroelectric properties in monolayer ferroelectric thin films has been preliminarily discussed [30-32], the synergistic mechanism of strain gradient and interface coupling in multilayer heterostructures (such as (PbTiO$_3$)$_n$/ (SrTiO$_3$)$_n$) has not been fully clarified, which is essential to optimize the electrical and mechanical properties of flexible ferroelectric heterojunction devices.

Therefore, the (SrTiO$_3$)$_{10}$/ (PbTiO$_3$)$_{10}$/ (SrTiO$_3$)$_{10}$ (STO/PTO/STO, where the subscript number represents the number of unit cells) flexible heterogeneous trilayer model is selected as the research object, and the domain structure evolution behavior of the system under different bending strain gradients is systematically discussed in this paper. By gradually adjusting the bending curvature radius, the evolution of ferroelectric domains under U-bending and N-bending strain modes is investigated, and the variation trend of the macroscopic hysteresis loop is evaluated using phase-field simulations. The results of this study reveal the mechanisms by which bending strain gradients affect the domain structures and polarization-switching behaviors of multilayer ferroelectric thin films, thereby further expanding the application potential of flexible mechanical-control strategies in heterogeneous ferroelectric architectures. This study not only deepens the understanding of the mechanical–electrical coupling mechanisms in multilayer ferroelectric thin films, but also provides a theoretical basis for the optimization of ferroelectric properties in future flexible ferroelectric devices.

## 2. Phase-field model

In this study, the polarization switching behavior of flexible ferroelectric trilayer films under bending deformation (U-type and N-type bending bending) is investigated by phase-field simulation. Simulation parameters are listed in **Table 1**[33,34]. In the phase-field method, the order parameter is the core variable to describe the microstructure evolution of the system. It is usually a continuous function varying with space and time, and is used to characterize the evolution behavior of different phases of materials during phase transformation [35-37]. In the ferroelectric phase-field model, the order parameter $P$ usually represents the spontaneous polarization of the material, which is essentially a three-dimensional vector field, $P = (P_x, P_y, P_z)$, where $P_x$, $P_y$, and $P_z$ represent the polarization components along the $x$, $y$, $z$ directions, respectively ($x$, $y$, $z$ three axes represent the [100], [010], and [001] crystal directions). The domain structure of ferroelectric materials is determined by the spatial distribution of polarized $P$, and the interaction and evolution of different domains are regulated by external stress, electric field, temperature and other factors. In particular, in flexible ferroelectric thin film systems, mechanical deformations such as applied bending can significantly affect the spatial distribution of polarization, resulting in the switching of domain structure. The dynamic evolution of the polarization vector is governed by a time-dependent phase field equation of the form:

$$\frac{\partial P_i(r,t)}{\partial t} = -L \frac{\delta F}{\delta P_i(r,t)}, i=1,2,3, \qquad (1)$$

where $r$ denotes the spatial coordinate of the system, $t$ denotes time, and $L$ is a kinetic coefficient related to the mobility of the domain wall. The evolution equation describes the relaxation process of ferroelectric polarization with time and finally reaches a stable thermodynamic equilibrium state. The $F$ represents the total free energy composed of Landau free energy, gradient energy, elastic energy, electrostatic energy and flexoelectric coupling energy:

$$F = \iiint_V [f_{Land}(P_i) + f_{grad}(P_{i,j}) + f_{elas}(P_i, \varepsilon_{ij}) + f_{ele}(P_i, E_i) + f_{flexo}(P_i, \varepsilon_{kl}, P_{i,j}, \varepsilon_{kl,j})] dV. \qquad (2)$$

Under the stress-free boundary condition, the Landau free energy density $f_{Land}$ of the thin film system can be expanded in a polynomial form through the components of the polarization vector as follows:

$$\begin{aligned} f_{Land}= & \alpha_1(P_x^2+P_y^2+P_z^2)+\alpha_{11}(P_x^4+P_y^4+P_z^4) \\ & +\alpha_{12}(P_x^2P_y^2+P_y^2P_z^2+P_z^2P_x^2) \\ & +\alpha_{111}(P_x^6+P_y^6+P_z^6) \\ & +\alpha_{112}[(P_x^4P_y^2+P_y^4P_x^2)+(P_y^4P_z^2+P_z^4P_y^2) \\ & +(P_x^4P_z^2+P_z^4P_x^2)]+\alpha_{123}P_x^2P_y^2P_z^2, \end{aligned} \quad (3)$$

where $\alpha_1$, $\alpha_{11}$, $\alpha_{12}$, $\alpha_{111}$, $\alpha_{112}$, and $\alpha_{123}$ are the material coefficients of the Landau free energy expansion, where $\alpha_1$ is temperature dependent $T$.

The gradient energy density $f_{grad}$ is related to the gradient of polarization and describes the contribution of the domain wall to the total free energy. It can be expressed as

$$\begin{aligned} f_{grad}= & \frac{1}{2}G_{11}(P_{x,x}^2+P_{y,y}^2+P_{z,z}^2) \\ & +G_{12}(P_{x,x}P_{y,y}+P_{y,y}P_{z,z}+P_{x,x}P_{z,z}) \\ & +\frac{1}{2}G_{44}[(P_{x,y}+P_{y,x})^2 \\ & +(P_{x,z}+P_{z,x})^2+(P_{y,z}+P_{z,y})^2] \\ & +\frac{1}{2}G'_{44}[(P_{x,y}-P_{y,x})^2 \\ & +(P_{x,z}-P_{z,x})^2+(P_{y,z}-P_{z,y})^2], \end{aligned} \quad (4)$$

where $G_{11}$, $G_{12}$, $G_{44}$, and $G'_{44}$ are the gradient energy coefficients. We adopt Einstein notation for tensors, with a comma in the subscript denoting spatial differentiation, such as $P_{i,j}=\partial P_i/\partial x_j$.

The elastic energy density $f_{elas}$ of the system is expressed as

$$\begin{aligned} f_{elas}= & \frac{1}{2}C_{11}(e_{xx}^2+e_{yy}^2+e_{zz}^2)+C_{12}(e_{xx}e_{yy} \\ & +e_{yy}e_{zz}+e_{zz}e_{xx})+2C_{44}(e_{xy}^2+e_{yz}^2+e_{xz}^2), \end{aligned} \quad (5)$$

where $C_{11}$, $C_{12}$, and $C_{44}$ are the elastic coefficients, and $e_{ij}$ represents the elastic strain. In most perovskite systems, the elastic strain is $e_{ij}=\varepsilon_{ij}-\varepsilon_{ij}^0$. While the eigenstrain $\varepsilon_{ij}^0$ is related to the spontaneous polarization of the material by

$$\varepsilon_{xx}^0 = Q_{11}P_x^2 + Q_{12}(P_y^2 + P_z^2),$$
$$\varepsilon_{yy}^0 = Q_{11}P_y^2 + Q_{12}(P_x^2 + P_z^2),$$
$$\varepsilon_{zz}^0 = Q_{11}P_z^2 + Q_{12}(P_y^2 + P_x^2),$$
$$\varepsilon_{xy}^0 = Q_{44}P_xP_y, \tag{6}$$
$$\varepsilon_{yz}^0 = Q_{44}P_yP_z,$$
$$\varepsilon_{xz}^0 = Q_{44}P_xP_z,$$

Where $Q_{11}$, $Q_{12}$, and $Q_{44}$ are the electrostrictive coefficients of the material.

The expression for the electrostatic energy density $f_{ele}$ is

$$f_{ele} = -\frac{1}{2}\varepsilon_0\kappa_b(E_x^2 + E_y^2 + E_z^2) - (E_xP_x + E_yP_y + E_zP_z), \tag{7}$$

where $E_x$, $E_y$, and $E_z$ are the electric field components; $\varepsilon_0$ is the dielectric constant of vacuum; $\kappa_b$ is the relative dielectric constant of the material. According to the theory of electrostatics, the built-in electric field can be represented by $E_i = -\varphi_{,i}$, where $\varphi$ is the electric potential. The distribution of the electric potential is solved by the electrostatic equilibrium equation:

$$\nabla \cdot \boldsymbol{D} = 0,$$
$$\nabla \cdot (\varepsilon_0 E_i + P_i) = 0. \tag{8}$$

In addition, if the strain gradient effect introduced by non-uniform strain is considered, the flexoelectric coupling energy $f_{flexo}$ of the system is introduced, which can be regarded as the coupling effect between strain gradient and polarization, and can be expressed as

$$f_{flexo} = \frac{1}{2}f_{ijkl}(P_{i,j}\varepsilon_{kl} - \varepsilon_{kl,j}P_i), (i,j,k,l=1,2,3), \tag{9}$$

where $f_{ijkl}$ is the flexoelectric coupling coefficient, which, for a cubic crystal, generally has three independent components: the longitudinal coupling coefficient $f_{11}$ ($=f_{1111}$), the transverse coupling coefficient $f_{12}$ ($=f_{1122}$), and the shear coupling coefficient $f_{44}$ ($=f_{1212}$).

Fig. 1 shows the phase-field calculation model of flexible STO/PTO/STO trilayer ferroelectric thin film under U-type (concave-up) and N-type (concave-down) bending deformation States. The simulation dimensions are $150\Delta x \times 150\Delta x \times 30\Delta x$, where $\Delta x = 0.4$ nm. The thickness of both PTO and STO layers are $10\Delta x$. In the simulation, the top of the film is set as a stress-free boundary, while the bottom surface is mechanically clamped to a flexible substrate, thereby more accurately capturing the mechanical constraint conditions encountered in experiments. To ensure the spatial continuity of the polarization and strain fields, periodic boundary conditions are applied along the in-plane directions of the computational domain, which eliminates artificial boundary truncation and allows for a more accurate description of the intrinsic domain-evolution characteristics. The open-circuit electrical boundary condition is used to calculate the domain structure of the STO/PTO/STO trilayer ferroelectric thin film in the bending state ($D_i n_i = 0$, $D_i$ is the electric displacement vector, $n_i$ refers to the component of the unit vector *n* solving the normal on the surface). To compute the hysteresis loops consistent with typical electrode configurations in experiments, short-circuit boundary conditions are adopted, in which the top and bottom electrode potentials are set to 0 V or an applied voltage. The polarization boundary conditions on both surfaces of the film are treated as free boundary conditions. Fig. 1(a) illustrates the trilayer structure and the corresponding bending radius $R$ under U-type and N-type bending. The radius of curvature $R$ determines the strain distribution throughout the system. As $R$ decreases (indicating an increase in bending severity), the stress state and polarization response within the film change significantly, thereby influencing the stability and switching dynamics of the ferroelectric domain structure. It is important to note that although both bending modes introduce strain gradients in the ferroelectric trilayer, their stress-distribution patterns differ markedly. Fig. 1(b) further presents the strain distribution in the ferroelectric heterostructure under U-type and N-type bending. The calculations show that, under U-type bending, the region below the neutral layer experiences in-plane tensile strain, whereas the region above the neutral layer is subjected to in-plane compressive strain. In contrast, for N-type bending, the region above the neutral layer is tensile while the region below is compressive. The formation of this nonuniform strain profile is primarily attributed to bending deformation; the direction of the strain gradient depends on the bending mode and exhibits mirror symmetry across different regions. During bending, the upper and lower layers of the ferroelectric heterostructure undergo different degrees of tension or compression, which induces a pronounced lattice-scale strain gradient. Because the flexible substrate directly bears the external load and exhibits the largest deformation near the substrate/film interface, the strain magnitude is most significant in this region. It is also noteworthy that the computational method and experimental design employed here are consistent with common bending configurations used in practical studies, where deformation of the flexible substrate drives the bending of the ferroelectric heterostructure [19,20,38], thereby governing the distribution of the strain field and its influence on polarization configurations at the macroscopic scale.

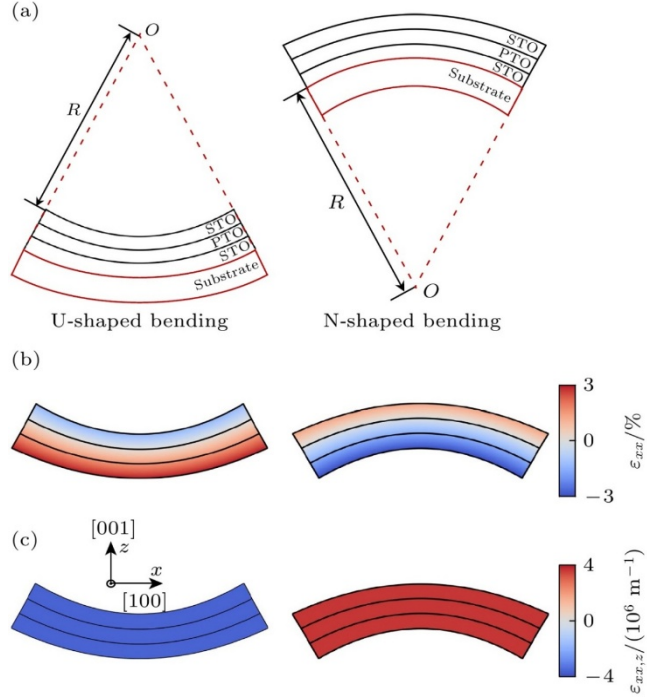

**Figure 1.** (a) Schematic diagrams of the two bending configurations of the ferroelectric trilayer film, where $R$ is the radius of curvature; (b) in-plane strain $\varepsilon_{xx}$ of the ferroelectric trilayer film under U-shaped and N-shaped bending; (c) strain gradient $\varepsilon_{xx,z}$ of the ferroelectric trilayer film under U-shaped and N-shaped bending.

**Table 1.** Material parameter values in the phase-field simulations (SI unit, $T$=300 K).

|  | Variable | Numerical value | Variable | Numerical value |
|---|---|---|---|---|
| PTO | $\alpha_1/(10^8$ J·m·C$^{-2})$ | $-1.706$ | $Q_{11}/(\text{m}^4\cdot\text{C}^{-2})$ | 0.089 |
|  | $\alpha_{11}/(10^7$ J·m$^5$·C$^{-4})$ | $-7.3$ | $Q_{12}/(\text{m}^4\cdot\text{C}^{-2})$ | $-0.026$ |
|  | $\alpha_{12}/(10^8$ J·m$^5$·C$^{-4})$ | 7.5 | $Q_{44}/(\text{m}^4\cdot\text{C}^{-2})$ | 0.0675 |
|  | $\alpha_{111}/(10^8$ J·m$^9$·C$^{-6})$ | 2.6 | $G_{11}/(10^{-10}$ N·m$^4$·C$^{-2})$ | 1.44 |
|  | $\alpha_{112}/(10^8$ J·m$^9$·C$^{-6})$ | 6.1 | $G_{12}/(\text{N}\cdot\text{m}^4\cdot\text{C}^{-2})$ | 0 |
|  | $\alpha_{123}/(10^8$ J·m$^9$·C$^{-6})$ | $-3.7$ | $G_{44}, G'_{44}/(10^{-11}$ N·m$^4$·C$^{-2})$ | 7.2 |
|  | $c_{11}/(10^{11}$ J·m$^{-3})$ | 2.3 | $f_{11}/\text{V}$ | 1.6 |
|  | $c_{12}/(10^{11}$ J·m$^{-3})$ | 1 | $f_{12}/\text{V}$ | $-0.8$ |
|  | $c_{44}/(10^{10}$ J·m$^{-3})$ | 7 | $f_{44}/\text{V}$ | 0.15 |
| STO | $\alpha_1/(10^8$ J·m·C$^{-2})$ | 2.017 | $Q_{44}/(\text{m}^4\cdot\text{C}^{-2})$ | 0.00957 |
|  | $\alpha_{11}/(10^9$ J·m$^5$·C$^{-4})$ | 1.7 | $G_{11}/(10^{-10}$ N·m$^4$·C$^{-2})$ | 1.44 |
|  | $\alpha_{12}/(10^9$ J·m$^5$·C$^{-4})$ | 4.45 | $G_{12}/(\text{N}\cdot\text{m}^4\cdot\text{C}^{-2})$ | 0 |
|  | $c_{11}/(10^{11}$ J·m$^{-3})$ | 3.3 | $G_{44}, G'_{44}/(10^{-11}$ N·m$^4$·C$^{-2})$ | 7.2 |
|  | $c_{12}/(10^{11}$ J·m$^{-3})$ | 1 | $f_{11}/\text{V}$ | $-3.21$ |
|  | $c_{44}/(10^{11}$ J·m$^{-3})$ | 1.25 | $f_{12}/\text{V}$ | 1.47 |
|  | $Q_{11}/(\text{m}^4\cdot\text{C}^{-2})$ | 0.0457 | $f_{44}/\text{V}$ | 1.07 |
|  | $Q_{12}/(\text{m}^4\cdot\text{C}^{-2})$ | $-0.0135$ | $\varepsilon_r$ (PTO/STO) | 20 |

# 3. Results and Discussion

Fig. 2 shows the evolution of the domain structure in STO/PTO/STO trilayer ferroelectric thin films under U-type bending deformation. In this system, a lattice-mismatch strain arises between the heterogeneous layers due to the different lattice parameters of PTO and STO (3.955 Å and 3.905 Å, respectively). Specifically, the PTO layer experiences in-plane compressive strain, whereas the STO layer is subjected to in-plane tensile strain. This interlayer strain mismatch exerts a significant influence on the polarization distribution, the stability of the domain structures, and the response of the system under external-field control.

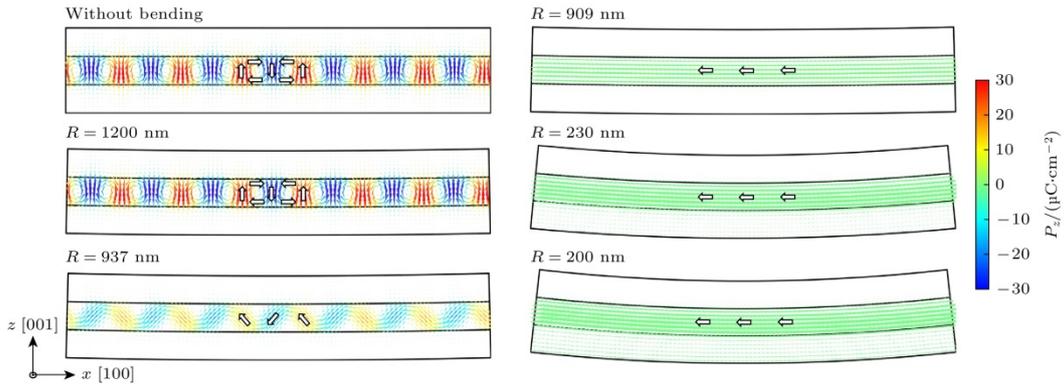

**Figure 2.** Domain evolution behavior in STO/PTO/STO trilayer films under different U-shaped bending radii.

In the initial state without bending deformation ($R \to \infty$), the ferroelectric domain structure in the PTO layer is characterized by polar-vortex pairs, which are closely associated with the competitive balance among the elastic interactions, electrostatic energy, and polarization-gradient energy of the system. The polarization vectors in the cores of these vortices rotate along closed trajectories, and the vortex-core diameter is approximately 4–5 nm, consistent with the polar-vortex structures observed by Yadav et al. [25]. The formation of the vortex state essentially represents an energy-minimization strategy through local polarization rotation, in which the electrostatic screening effect at the ferroelectric/dielectric interface plays a crucial role in stabilizing the topology. With the decrease of the bending curvature radius $R$ (that is, the bending degree increases), the strain state of the PTO layer gradually changes, which affects the stability and topology of the domain structure. When $R > 937$ nm, the system still maintains the original polar vortex pair characteristics, indicating that the accumulation of interfacial strain energy is not enough to drive the topological phase transition of the domain structure under this strain condition. However, when the $R$ decreases to 937 nm, the polarization configuration in the PTO layer changes suddenly, and the original polar vortex pairs evolve into end-to-end vortex States, and the spatial distribution shows a zigzag arrangement. This structural reconstruction can be attributed to the strain state introduced by bending deformation, which not only regulates the local polarization rotation direction, but also significantly changes the energy distribution of the system, making the original vortex structure no longer stable. When the $R$ is further reduced ($R = 200$ nm), the top surface of the STO/PTO/STO trilayer is mainly subjected to compressive strain ($\varepsilon_{xx} \approx -1.35\%$), while the bottom surface is mainly subjected to tensile strain ($\varepsilon_{xx} \approx 3\%$). At this time, the in-plane tensile strain of the system is dominant, which eventually leads to the complete transformation of the polarization configuration in the PTO layer into the in-plane $a$ domain.

The Fig. 3 shows the domain structure evolution in STO/PTO/STO trilayer ferroelectric thin films under N-type bending deformation. Compared with U-bending, the primary feature of N-type bending is that the deformation boundary condition is applied to the bottom layer (flexible substrate). This leads to compressive strain dominating the lower part of the system, which strongly influences the local polarization configuration and may induce new domain-evolution pathways.

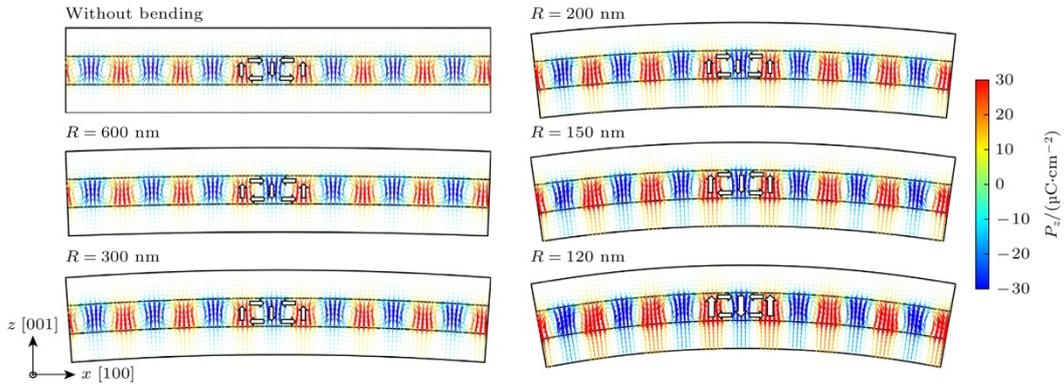

**Figure 3.** Domain evolution behavior in STO/PTO/STO trilayer films under different N-shaped bending radii.

When the bending radius is large (i.e., the bending degree is small), the polar vortex pairs in the PTO layer are still stable, indicating that the strain gradient in the system is insufficient to disrupt the energy balance of the original vortex state. However, as the radius of curvature $R$ of the N-type bend decreases further (i.e., the degree of bend increases). For example, when $R = 120$ nm, the N-type bending deformation leads to the tensile strain on the upper surface ($\varepsilon_{xx} \approx 2.27\%$) and the compressive strain on the lower surface ($\varepsilon_{xx} \approx -5\%$), and the in-plane compressive strain of the system is further enhanced, resulting in the gradual reduction of the size of the polar vortex pair. This phenomenon indicates that the bending-induced strain regulation effect can effectively suppress the spatial expansion of the vortex core and promote the local polarization vector to gradually align in the out-of-plane direction. When the bending degree is further increased, the polarization configuration of the system changes obviously: the polar vortex pair gradually evolves into an out-of-plane $c$ domain with a 180-degree domain wall, that is, the polarization direction is aligned along the normal direction of the film. This domain switching behavior can be attributed to two factors: one is that the inhomogeneous compressive strain introduced by bending deformation suppresses the polarization rotation degree of freedom, which reduces the in-plane polarization component and dominates the out-of-plane polarization; Secondly, the dielectric control effect of the STO layer is enhanced, and the ferroelectric polarization in the STO layer gradually appears under the strong N-type bending strain, which further promotes the formation and stability of the 180 ° domain wall in the PTO layer. In addition, as the degree of bending is further increased, the polarization features of the STO layer itself become significant, eventually forming out-of-plane *c* domains with 180 ° domain walls. Under N-type bending, the change of strain gradient direction not only affects the stability of polar vortex pairs, but also plays a significant role in regulating the spatial distribution of ferroelectric domain structure. The evolution behavior of domain structure regulated by related bending strain shows that the bending direction of U-type or N-type can not only drive the topological phase transition of ferroelectric domains, but also selectively stabilize specific polarization structure by regulating the direction of strain gradient.

After studying the effect of bending deformation on the evolution of domain structure in STO/PTO/STO trilayer, the control mechanism of bending deformation on the macroscopic hysteresis loop of trilayer was further explored. As shown in the Fig. 4, the hysteresis loop of the STO/PTO/STO trilayer film under U-bend condition shows a significant overall shift. It is worth pointing out that although both U-type and N-type bending may introduce a certain degree of strain in the PTO layer, the more important thing is that they form strain gradients in different directions in the thickness direction, and the strain gradient $\varepsilon_{xx,z}$ in the system will lead to the generation of transverse flexural electric field $E_z^{\text{flexo}}=f_{12}\varepsilon_{xx,z}$. In general, the flexoelectric coupling coefficient $f_{12}$ of perovskite materials is positive, which means that different bending morphologies (U-type or N-type) will lead to opposite flexoelectric fields. Under U-bending ($R = 240$ nm), the top surface of the STO/PTO/STO trilayer is mainly subjected to compressive strain ($\varepsilon_{xx} \approx -1.13\%$) and the bottom surface is mainly subjected to tensile strain ($\varepsilon_{xx} \approx 2.5\%$), resulting in a negative strain gradient in the thickness direction ($\varepsilon_{xx,z} \approx -3.03 \times 10^6$ m$^{-1}$), which induces a negative deflection electric field in the out-of-plane direction. The flexural electric field effectively regulates the polarization reversal process, causing the hysteresis loop to shift to the lower right as a whole (as shown in the enlarged view of the central region of Fig. 4).

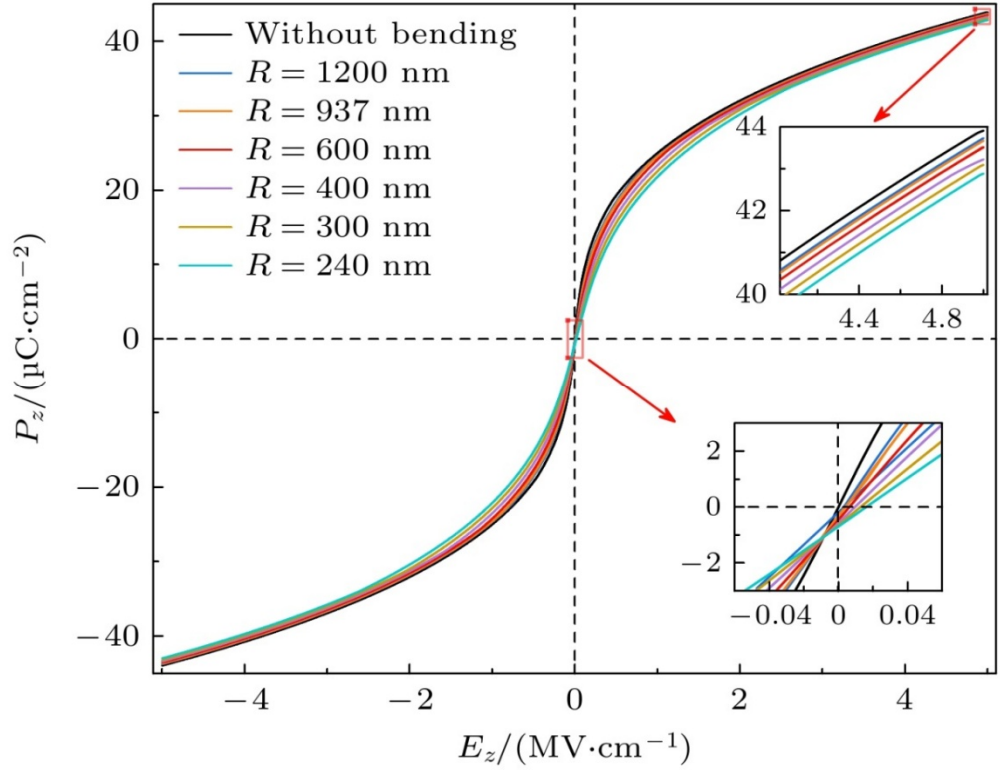

**Figure 4.** Hysteresis loops of STO/PTO/STO trilayer films under different U-shaped bending radii.

Further analysis shows that with the increase of the degree of U-bending (that is, the bending curvature radius $R$ becomes smaller), this shift effect gradually increases, which leads to the change of coercive electric field ($E_c$), maximum polarization ($P_{max}$) and remanent polarization ($P_r$), as listed in Table 2. In the unbent state, the coercive electric field and the remanent polarization are close to zero because the inside of the STO/PTO/STO trilayer film is mainly a polar vortex pair structure and an in-plane single domain is formed in the zero electric field state after a cyclic electric field is applied. In contrast, when the trilayer is in U-bend and the radius of curvature $R$ = 240 nm, the coercive field increases from 0 kV/cm to 16.67 kV/cm, while the remanent polarization becomes negative –0.69 μC/cm². However, due to the in-plane tensile strain induced by U-bending, the electric dipoles tend to arrange in the in-plane direction, which leads to the decrease of the overall maximum polarization. The maximum polarization decreases from 43.90 μC/cm² in the unbent state to 42.86 μC/cm². The overall shift of the hysteresis loop is closely related to the evolution of the domain structure, indicating that bending deformation can not only control the domain structure through strain, but also significantly change the macroscopic ferroelectric response of the system through the flexoelectric effect.

**Table 2.** Coercive electric field, maximum polarization, and remnant polarization of the ferroelectric hysteresis loop in STO/PTO/STO trilayer films under U-shaped bending deformation.

| U-bend-$R$/nm | $\varepsilon_{xx,z}$/($10^6$ m$^{-1}$) | $E_c$/(kV·cm$^{-1}$) | $P_{max}$/(μC·cm$^{-2}$) | $P_r$/(μC·cm$^{-2}$) |
|---|---|---|---|---|
| Unbent | 0 | 0 | 43.90 | 0 |
| 1200 | –0.61 | 2.78 | 43.72 | –0.29 |
| 937 | –0.77 | 2.78 | 43.64 | –0.31 |
| 600 | –1.14 | 5.56 | 43.52 | –0.46 |
| 400 | –1.82 | 11.11 | 43.30 | –0.57 |
| 300 | –2.42 | 13.89 | 43.05 | –0.64 |
| 240 | –3.03 | 16.67 | 42.86 | –0.69 |

The Fig. 5 shows the evolution of the polarization distribution of STO/PTO/STO trilayer under U-bend deformation under external electric field. In the unbent state and without an applied electric field (0 MV/cm), the electric dipoles inside the PTO layer spontaneously form a periodic array of clockwise-counterclockwise polar vortex pairs. When a forward electric field is applied, the polarization distribution is significantly reconstructed. For example, when the electric field increases to 0.21 MV/cm, the vortex core is gradually annihilated, and part of the out-of-plane polarization component is deflected along the direction of the applied electric field and gradually turns outward and upward. Finally, the out-of-plane single domain structure is formed at an electric field of 5 MV/cm, and the direction of all electric dipoles in the PTO layer is consistent with the direction of the applied electric field. It is worth noting that when the electric field is reduced from the out-of-plane to 0, the system no longer returns to the original vortex state, but becomes an in-plane single domain, which indicates that the system has a certain hysteresis effect. When the trilayer is in U-bending state (radius of curvature $R$ = 240 nm), the PTO layer is subjected to a dominant in-plane tensile strain. Due to the introduction of tensile strain, the stability of the out-of-plane polarization component ($c$ domain) is reduced, and the polarization vector tends to align along the in-plane direction, thus forming an in-plane single-domain structure. When the electric field increases to 5 kV/cm, the polarization distribution tends to be uniform and the system enters the in-plane monodomain state.

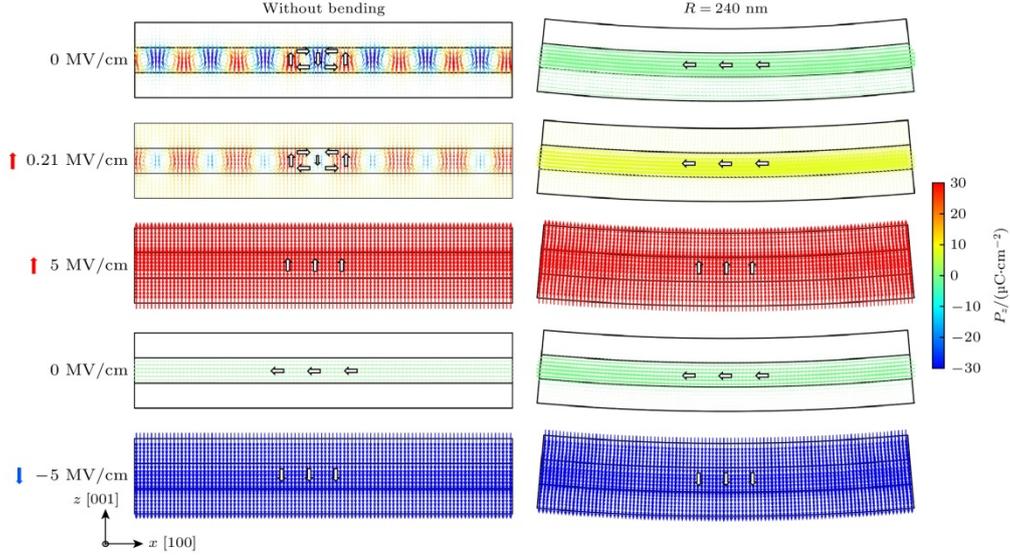

**Figure 5.** Electric field-modulated polarization distribution in STO/PTO/STO trilayers under U-shaped bending deformation.

Next, we study the macroscopic hysteresis loop behavior of the flexible trilayer under N-type bending deformation, as shown by Fig. 6. Under the condition of N-type bending, the STO/PTO/STO trilayer film generates a positive flexural electric field in the out-of-plane direction, which leads to the overall upper left shift of the electric hysteresis loop (as shown in the enlarged image of the central region of the Fig. 6), and with the increase of the degree of N-type bending (that is, the radius of curvature $R$ gradually decreases), the shift gradually increases, indicating that the control effect of the flexural electric field is significantly enhanced with the increase of bending deformation. As a result, the coercive electric field, the maximum polarization and the remanent polarization of the three-layer film are all changed. See Table 3. In the unbent state, the coercive electric field of the system is close to zero because the PTO layer is mainly composed of vortex pairs; When the radius of curvature $R = 200$ nm, the coercive electric field changes to –19.44 kV/cm. In terms of remanent polarization, the $P_r$ in the unbent state are approximately zero, which is due to the fact that the system is mainly in the vortex state at zero field, and the polarization has no obvious spontaneous orientation; When $R = 200$ nm, the $P_r$ increases to 1.42 μC/cm², indicating that the N-type bend-induced forward flexure electric field promotes the formation of the out-of-plane polarization component, making it easier for the system to maintain the remanent polarization at zero field. In terms of the maximum polarization, the in-plane compressive strain introduced by the N-type bending enhances the tendency of the electric dipoles to align in the out-of-plane direction, thus increasing the saturation polarization of the system; $P_{max}$ is 43.90 μC/cm² in the unbent state, while $P_{max}$ increases to 45.15 μC/cm² when $R = 200$ nm, indicating that N-type bending contributes to the enhancement of ferroelectric response. The physical mechanism of this phenomenon is mainly attributed to the bending-induced strain gradient control. The positive flexure electric field generated by N-type bending promotes the alignment of the out-of-plane polarization direction, which enhances the residual polarization at zero field and causes the electric hysteresis loop to shift to the upper left; At the same time, the in-plane compressive strain effect further enhances the stability of the out-of-plane polarization, increases the maximum polarization, and enhances the critical electric field required for polarization switching, resulting in a significant increase in the coercive electric field. To sum up, the residual bending not only affects the overall shift of the ferroelectric hysteresis loop, but also changes the polarization switching characteristics of the system through strain control and flexural electric field effect, which provides an important physical insight into the mechanical-electric coupling mechanism of flexible ferroelectric thin films.

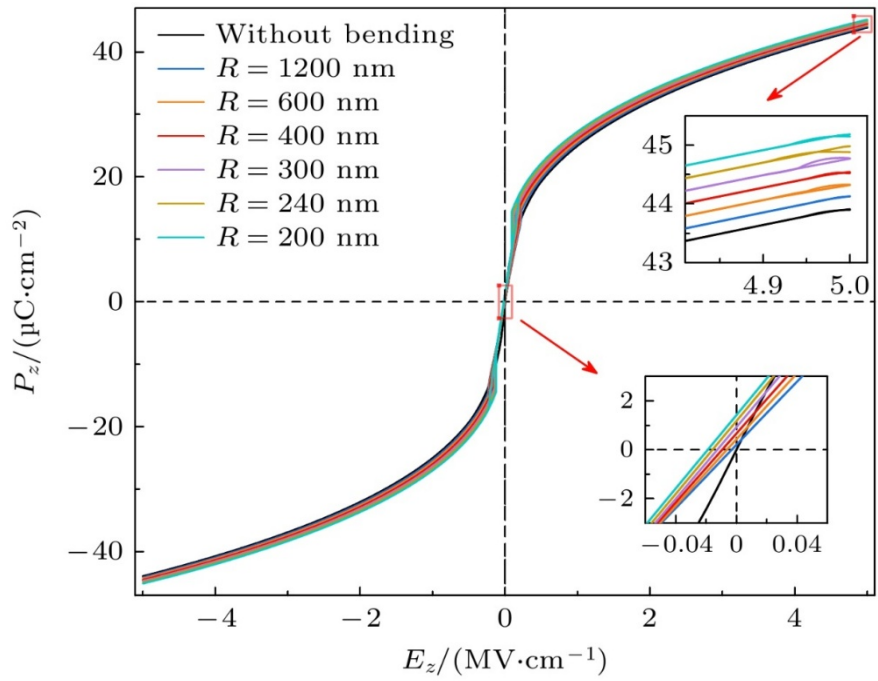

**Figure 6.** Hysteresis loops of STO/PTO/STO trilayer films under different N-shaped bending radii.

**Table 3.** Coercive electric field, maximum polarization, and remnant polarization of the ferroelectric hysteresis loop in STO/PTO/STO trilayer films under N-shaped bending deformation.

| N-type bend-$R$/nm | $\varepsilon_{xx,z}/(10^6 \text{ m}^{-1})$ | $E_c/(\text{kV·cm}^{-1})$ | $P_{max}/(\mu\text{C·cm}^{-2})$ | $P_r/(\mu\text{C·cm}^{-2})$ |
|---|---|---|---|---|
| Unbent | 0 | 0 | 43.90 | 0 |
| 1200 | 0.61 | −2.78 | 44.13 | 0.21 |
| 600 | 1.14 | −5.56 | 44.33 | 0.43 |
| 400 | 1.82 | −8.33 | 44.53 | 0.67 |
| 300 | 2.42 | −13.89 | 44.78 | 0.93 |
| 240 | 3.03 | −16.67 | 44.88 | 1.18 |
| 200 | 3.61 | −19.44 | 45.15 | 1.42 |

The Fig. 7 shows the dynamic evolution of the polarization distribution of STO/PTO/STO trilayer with electric field under N-type bending deformation. When the radius of curvature $R$ of N-type bending is 200 nm and no electric field (0 MV/cm) is applied, the tensile strain ($\varepsilon_{xx} \approx 1.5\%$) on the upper surface and the compressive strain ($\varepsilon_{xx} \approx -3\%$) on the lower surface caused by bending deformation induce an out-of-plane forward strain gradient field ($\varepsilon_{xx,z} \approx 3.61 \times 10^6$ m$^{-1}$). Due to the dominant effect of compressive strain in the whole system, the polarization configuration in the PTO layer is mainly a periodic vortex-antivortex array. However, compared with the U-bending, the in-plane compressive strain introduced by the N-bending makes the lattice shrink along the out-of-plane direction, which further stabilizes the out-of-plane $c$ domain and significantly inhibits the nucleation process of the in-plane $a$ domain. When the applied out-of-plane electric field increases to 0.21 MV/cm, the domain structure of the system has completely transformed into the out-of-plane $c$ domain, indicating that the flexure electric field control under N-type bending is enough to drive the polarization vector to align in the out-of-plane direction. As the electric field continues to increase to the forward saturation value ($E = 5$ MV/cm), the system enters a stable out-of-plane monodomain state, and its maximum polarization ($P_{max} = 45.15$ μC/cm²) is 2.9% higher than that of the unbent state (43.90 μC/cm²). This polarization enhancement effect is due to the synergistic effect between the strain gradient generated by N-type bending and the applied electric field, that is, the direction of the bending electric field is consistent with the direction of the applied electric field, which together drive the polarization vector to arrange along the out-of-plane direction and improve the polarization response ability of the system. After removing the external electric field, the remanent polarization of the system remains positive, which corresponds to the upper left shift of the hysteresis loop in Fig. 6, further verifying the asymmetric control of the strain gradient sign on the polarization reversal path. The dynamic process of polarization evolution revealed by Fig. 7 not only deepens the understanding of the mechanical-electrical synergistic effect in flexible ferroelectric thin films, but also provides a new theoretical basis for the performance optimization of flexible ferroelectric devices (such as multi-state storage, chiral sensing).

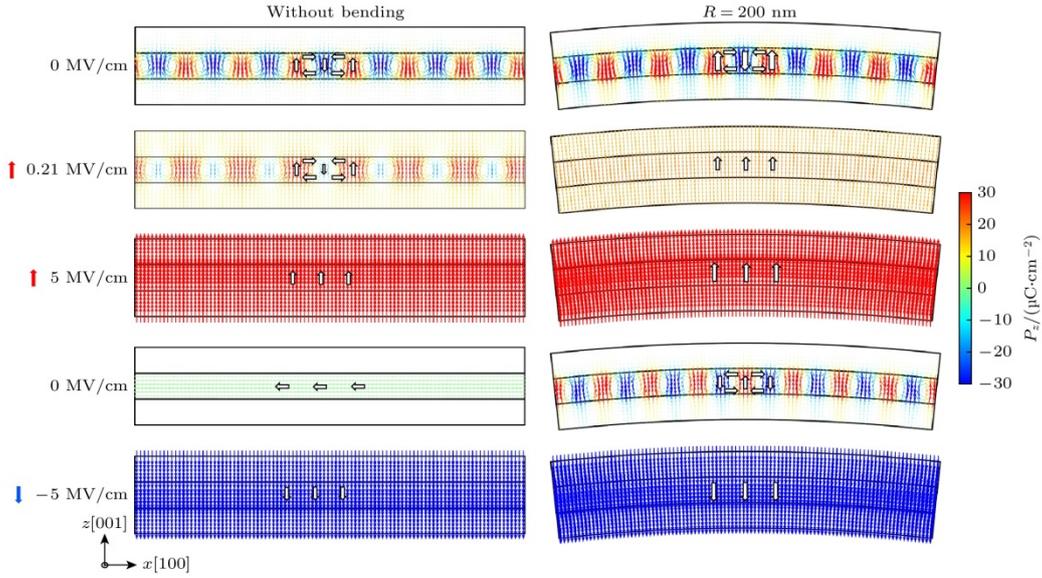

**Figure 7.** Electric field-modulated polarization distribution in STO/PTO/STO trilayer under N-shaped bending deformation.

## 4. Conclusion

In this paper, phase-field simulations are employed to explore the control mechanism of bending-induced strain gradients on the domain structure and macroscopic hysteresis behavior of STO/PTO/STO trilayer ferroelectric thin films. The results indicate that U-type and N-type bending deformations introduce strain gradients of opposite directions, thereby significantly altering the polarization distribution and the evolution of topological domains in the film. Under U-type bending, the induced negative flexoelectric field causes the hysteresis loop to shift toward the lower-right region. When the U-bend curvature radius is 240 nm, the coercive electric field increases to 16.67 kV/cm, the remanent polarization becomes negative (–0.69 μC/cm²), and the maximum polarization decreases to 42.86 μC/cm², indicating that the dominant in-plane tensile strain suppresses the stability of the out-of-plane polarization component. Conversely, the positive flexoelectric field generated by N-type bending causes the hysteresis loop to shift toward the upper-left region. When the N-type bending curvature radius is 200 nm, the coercive field changes to –19.44 kV/cm, the remanent polarization increases to 1.42 μC/cm², and the maximum polarization increases to 45.15 μC/cm², reflecting the enhancement of out-of-plane polarization under dominant in-plane compressive strain. In addition, the bending-induced strain gradient drives a topological phase transition from polar-vortex pairs to either in-plane or out-of-plane single-domain states, revealing the synergistic interplay between the strain gradient and interfacial electrostatic coupling. These findings not only clarify the mechanisms by which bending deformation regulates ferroelectric domain structures and electrical responses, but also provide a theoretical foundation for the design and optimization of flexible ferroelectric devices.

We thank Dr. LI Yongheng, Department of Mechanics, Beijing Institute of Technology, for discussions.